\documentclass[prb,amsmath,amssymb,preprint]{revtex4}

\usepackage{graphicx}
\usepackage{bm}

\begin{document}

\title{Three weird facts about quantum mechanics: 
What Bohr, Schr\"{o}dinger, and Einstein actually said}
\author{M. A. Reynolds}
\email{anthony.reynolds@erau.edu}
\affiliation{Department of Physical Sciences,
Embry-Riddle Aeronautical University,
Daytona Beach, Florida, 32114}

\date{\today}

\begin{abstract}
The procedure used to ``do physics'' in the macroscopic world is familiar:
You take an object, 
start it off with a particular position and velocity,
subject it to known forces (say gravity or friction, or both),
and follow its trajectory.
You then measure the dynamical properties (say position or energy) of that object at a later time
and compare those measurements with the prediction using Isaac Newton's laws of motion.
Newton's laws directly predict what those quantities should be at that later time,
so the comparison is straightforward.
However, the microscopic laws of physics, 
quantum mechanics,
aren't so simple.
The quantum concepts are so alien and counterintuitive that the language used to describe the
mathematics and physics is often ambiguous and confusing.
Therefore, for you to \emph{really} understand what is going on,
some mathematics must be used.
This is my attempt to explain some ``weird facts'' using
a minimum of mathematics,
yet eliminating as much ambiguity as possible.
\end{abstract}

\maketitle

\section{Niels Bohr 1913}

One of the first weird facts about quantum mechanics came out of Niels Bohr's attempt to understand the atom
in 1913.
Only two years before, in 1911, Ernest Rutherford (with the help of Hans Geiger and Ernest Marsden) had shown
that atoms were composed of a tiny, massive nucleus with positive charge surrounded by electrons of negative charge.
Hydrogen, with only one electron, was the simplest of these.
Bohr had joined Rutherford's group in Manchester after receiving his doctorate from the University of Copenhagen
in 1911.
While there, he tried to understand the spectrum of hydrogen using Rutherford's ``planetary'' model as a starting point.

From the classical electrodynamics developed in the late 1800s,
it was known that the combination of a single electron orbiting a proton
(as the hydrogen nucleus was called) in planetary fashion was not stable.
The accelerating electron radiated electromagnetic waves, losing energy,
and was predicted to spiral into the proton in about $10^{-11}$ s.
Of course, as there is plenty of hydrogen around, this prediction must be wrong.
Bohr therefore hypothesized that the electron could only occupy a ``stationary state'' in which
it did \textit{not} spiral into the proton.
(What this stationary state was, or how it remained stationary, was left unexplained.)
He chose what would later be called ``energy eigenstates'' in order to correctly predict the spectrum
of hydrogen.\footnote{The details of his calculation are not
important for our purposes here, only the implications of the stationary state hypothesis.}

In modern notation,
a single particle in a stationary state 
(in this example, an electron in a hydrogen atom)
can be written
\begin{equation}\label{eq:StationaryState}
| \psi \rangle = | \psi_n \rangle .
\end{equation}
That is, the state of the electron, $| \psi \rangle$,
is a stationary state given by the quantum number $n$, $| \psi_n \rangle$,
and has energy
\begin{equation}\label{eq:HydrogenSpectrum}
E_n = \frac{-13.6 \textrm{ eV}}{n^2} ,
\end{equation}
where $n$ can take on the discrete integer values 1, 2, $\ldots$, $\infty$.
The 
``ket'' notation $| \psi \rangle$ is used to remind you that we are not really talking about a function,
but simply labeling the state that the electron is in.\footnote{This notation, $| \psi \rangle$, was invented by Paul Dirac in the 1920s.}
The stationary states (which will turn out to be solutions to Schr\"{o}dinger's equation,
described in the next section) 
can also be written simply as 
$| \psi \rangle =| n \rangle$.\footnote{There are really four quantum numbers that completely describe the state of the electron in a
hydrogen atom,
$n$, $\ell$, $m_\ell$, and $m_s$,
but for our purposes, all the weird physics can be understood with only one quantum number.}

The hydrogen spectrum that Bohr was trying to understand can be
obtained by assuming that the electron can ``jump'' from one stationary state to 
another stationary state (with a lower energy).
In this process,
energy is conserved and therefore the atom must emit a photon with an energy
equal to the energy difference between the two states
\begin{equation}
h\nu = \Delta E = E_m - E_n ,
\end{equation}
where $\nu$ is the frequency of the emitted photon and $h$ is Planck's constant.
This assumes that light comes in discrete units called ``quanta'' 
(now called a photon)
and each has energy $h\nu$
(an idea that was put forth by Max Planck in 1900 and used by Albert Einstein in 1905).
Bohr had therefore combined a previous idea (the quantum of light) with his own hypothesis (stationary states)
along with standard physics (energy conservation) and was able to predict the hydrogen spectrum perfectly!
However, physicists were still confused about what was \textit{meant} by a state.
And how did the electron know when to jump, and to what other state to jump to?
Why were these the only states that were stationary?
These are questions that we still have today,
but we answer them in a probabilistic fashion.
As physicist Abraham Pais put it
\begin{quote}
At a moment which cannot be predicted an excited atom makes a transition to its ground state by emitting a photon.
    Where was the photon before that time? It was not anywhere; it was created in the act of transition....
Is there a theoretical framework for describing how particles are made and how they vanish?
    There is: quantum field theory. It is a language, a technique, for calculating the
    probabilities of creation, annihilation, scatterings of all sorts of particles: photons,
    electrons, positrons, protons, mesons, others ...\footnote{A. Pais,
    \textit{Inward Bound} (Oxford University Press, New York), pp.\ 324-5.}
\end{quote}

\paragraph*{\underline{Weird Fact \#1}}
This, then, is our first weird fact.
In certain situations
electrons (particles, in general) occupy stationary states which have discrete energies
(completely contrary to the predictions of classical mechanics in which particles can have any energy)
and they emit or absorb photons when they make transitions between these states.
They are not allowed to have any other energy.

\section{Erwin Schr\"{o}dinger 1926}

The second weird fact about quantum mechanics arose with Erwin Schr\"{o}dinger's equation
\begin{equation}
H | \psi_n \rangle = E_n | \psi_n \rangle ,
\end{equation}
which he obtained by making some assumptions about the mathematical form of the state $|\psi \rangle$
and invoking energy conservation.
It looks very complicated
(for example, $H$ is something called the Hamiltonian operator), but for our purposes, all we need to know is that it
is what mathematicians call an eigenvalue equation.
What does this mean?
It means that there are solutions only for discrete values of the energy $E_n$,
which is called the eigenvalue.\footnote{In this context,
$|\psi_n \rangle$ is called the ``eigenfunction,'' or ``wave function.''}
This is just what Bohr had postulated!
In fact,
when Schr\"{o}dinger applied his equation to the hydrogen atom,
he was able to derive Bohr's result,
and Bohr's energies --- see Equation (\ref{eq:HydrogenSpectrum}) --- turned out to be just the energy eigenvalues.

Since this is a differential equation,
there are many possible solutions, one for each value of $n$ in the set $n = 1$, 2, 3, $\ldots$, $\infty$.
As is true for all linear differential equations,
the most general solution is a combination of all viable solutions,
and this means that the electron in a hydrogen atom doesn't have be in just \textit{one} state,
as assumed in Equation (\ref{eq:StationaryState}),
but can be in a ``superposition state''
\begin{equation}\label{eq:SuperState}
| \psi \rangle = c_1 |\psi_1 \rangle + c_2 |\psi_2 \rangle + c_3 |\psi_3 \rangle + \cdots
= \sum_n c_n | \psi_n \rangle ,
\end{equation}
that is, many states at once.
(This, of course, is counter to our everyday experience in which we never find objects in superposition states,
but always find them in a single state.)

\paragraph*{\underline{Weird Fact \#2}}
Again, contrary to the predictions of classical physics,
and counter to our everyday experience,
electrons (particles in general) are allowed to be in superposition states,
in which they can take on the characteristics of each of the stationary states (for example, the energy)
with a certain probability.

\subsection*{The measurement process}

This weirdness manifests itself when we talk about taking measurements of a system.
If an electron is in such a superposition state as depicted by Equation (\ref{eq:SuperState}),
quantum mechanics tells us that if you measure the energy,
you won't obtain just \textit{any} value for the result of your measurement,
but the only possible values will be $E_1$ or $E_2$ or $E_3$, etc.
In addition,
the probability of measuring a certain energy, say $E_n$, is given by the coefficient squared,
$|c_n|^2$.
Since the probability of measuring \textit{any} energy must be unity,
there must be a restriction $\sum_n |c_n|^2 = 1$.
This is called ``normalization.''
The way it is usually described is as follows.
When measuring the energy of one electron, there is no way of knowing which energy you will obtain,
but after measuring the energies of many \textit{identically prepared} electrons 
(this set of systems is called an ensemble),
the probability of obtaining the different energies will be as described.

Even though Schr\"{o}dinger had developed the equation that led to this formalism,
he thought that the situation was ridiculous.
To show this, he came up with a thought experiment (involving the infamous cat) to highlight how strange this is.
In his own words:
\begin{quote}
One can even set up quite ridiculous cases. A cat is penned up in a steel chamber, along with the following device (which must be secured against direct interference by the cat): in a Geiger counter there is a tiny bit of radioactive substance, so small, that perhaps in the course of the hour one of the atoms decays, but also, with equal probability, perhaps none; if it happens, the counter tube discharges and through a relay releases a hammer which shatters a small flask of hydrocyanic acid. If one has left this entire system to itself for an hour, one would say that the cat still lives if meanwhile no atom has decayed.\footnote{This quote is from a three-part paper,
E. Schr\"{o}dinger, ``Die gegenw\"{a}rtige Situation in der Quantenmechanik,'' \textit{Naturwissenschaften}, \textbf{23} 807-812; 823-828; 844-849 (1935). 
Translated by John D. Trimmer.}
\end{quote}
That is,
imagine a cat in an opaque box.
Also in this box is a vial of poison gas, and if this vial breaks, the cat will die.
Also in this box is a radioactive atom, arranged such that if the atom decays, 
it will trigger a small hammer to break the vial, killing the cat.
Now then, after one hour,
the cat is in a superposition state
(in this case there are only two states, i.e., $n=1$, 2, or more descriptively, ``alive'' or ``dead''),
\begin{equation}
|\psi_{cat} \rangle = c_{alive} |\psi_{alive} \rangle + c_{dead} | \psi_{dead} \rangle ,
\end{equation}
where we have chosen the radioactive substance so that the values of $c_{dead}$ and $c_{alive}$
are each equal to $\frac{1}{\sqrt{2}}$,
which means that the probabilities of the cat being alive or dead are each
\[
\left| c_{alive} \right|^2 = \left| c_{dead} \right|^2 = \left| \frac{1}{\sqrt{2}} \right|^2 = \frac{1}{2} .
\]

Now, even though Schr\"{o}dinger --- and many others --- thought this was weird, 
it is an observational fact that these rules predict accurately the outcomes of subatomic experiments.
That is,
Weird Fact \#2
(the possibility of particles being in states of superposition)
is an accurate description of the world,
as long as we interpret these states in the probabilistic fashion just described.

\section{Entanglement}

Our third weird fact comes into play when we consider two identical particles simultaneously,
for example, the two electrons in a helium atom.
Following our previous rules, you might think that each electron must occupy a stationary state,
or perhaps occupy a superposition state. 
In general, this is true, 
but the presence of one electron affects the other electron --- specifically,
one electron modifies the electric force experienced by the second electron, so the stationary states
(and their energies) are modified from the one-electron case.

But more important, 
there really is only \emph{one} wave function, or \emph{one} state, 
but it depends on the properties of \emph{both} electrons.
That is,
we can write the wave function of the two electrons as a ``product'' state
\begin{equation}
\label{eq:product}
| \psi \rangle = | \psi_n^a \rangle | \psi_m^b \rangle,
\end{equation}
which means that electron $a$ is in state $n$ and electron $b$ is in state $m$.
However,
in quantum mechanics, identical particles are truly identical.
In classical physics,
two white billiard balls, 
while they look the same,
can be distinguished by looking closely at any possible scuffs or scratches.
But two electrons, for example, are \emph{indistinguishable},
and no one, not even God, can tell them apart.
This means that we don't know whether 
electron $a$ is in state $n$ and electron $b$ is in state $m$ 
or vice versa.
Therefore, Equation (\ref{eq:product}) is not an accurate representation of the state of the system.
We must allow for the possibility that the particles switch places,
i.e., electron $a$ is in state $m$ and electron $b$ is in state $n$.
This means that the wave function must be written as 
\begin{equation}
\label{eq:entangled}
| \psi \rangle = | \psi_n^a \rangle | \psi_m^b \rangle 
+ | \psi_m^a \rangle | \psi_n^b \rangle ,
\end{equation}
which is called an ``entangled'' state.
The particles are entangled because we don't know which particle is in which state,
although we \emph{do} know that if one of them is in state $a$,
then the other \emph{must} be in state $b$.

In reality,
the wave function must be normalized,
and the state must be
\begin{equation}
\label{eq:entangled}
| \psi \rangle = \alpha | \psi_n^a \rangle | \psi_m^b \rangle 
+ \beta | \psi_m^a \rangle | \psi_n^b \rangle ,
\end{equation}
where $|\alpha|^2$ is the probability of finding electron $a$ in state $n$ and electron $b$ in state $m$
and $|\beta|^2$ is the probability of finding electron $a$ in state $m$ and electron $b$ in state $n$.
Of course,
normalization requires that 
$|\alpha|^2 + |\beta|^2 = 1$.
And since either case is equally probable, 
$|\alpha|^2 = |\beta|^2 = \frac{1}{2}$.
Mathematically,
there are two equally correct choices:
either 
$\alpha = \frac{1}{\sqrt{2}}$ and $\beta = \frac{1}{\sqrt{2}}$,
in which case $| \psi \rangle$ is called a symmetric wave function,
or
$\alpha = \frac{1}{\sqrt{2}}$ and $\beta = -\frac{1}{\sqrt{2}}$,
in which case $| \psi \rangle$ is called an anti-symmetric wave function.\footnote{It turns out that these two
choices describe two different kinds of particles.
Symmetric wave functions describe ``bosons,''
and 
anti-symmetric wave functions describe ``fermions.''}

\paragraph*{\underline{Weird Fact \#3}}
When the system consists of two identical (i.e., indistinguishable) particles,
they are in a superposition state together,
which is called an entangled state.
It is not known which particle is in which state,
but it is equally likely to be either.

\subsection*{Einstein 1935}

Einstein,
among others,
thought that this was weird (just like Schr\"{o}dinger had thought Weird Fact \#2 was weird),
and he spent many years arguing with Bohr about the meaning of superposition states and entangled states.
Much has been written about these debates,
and I can't do justice to them here.
However, I will discuss one aspect of the problem that philosophers of physics must wrestle with,
and which Einstein got wrong.\footnote{Modern writers make a big deal whenever they discover that
Einstein got something wrong,
but in this case he made a perfectly reasonable choice that was only shown to be wrong by 
experiments performed in the 1970s.}

Einstein was a realist.
This means that while he believed that the mathematical prescription of a superposition state
correctly predicted the probabilities of certain measurements
(as described above),
he thought that such a superposition did not completely describe the state of the object.
That is,
an electron really \emph{does} have a specific energy,
but we just don't know what it is.
Our knowledge is incomplete, and a deeper, more fundamental theory than quantum mechanics would tell us
exactly what the energy is.
Another way of thinking about it is that the standard quantum mechanical description claims that the
electron doesn't have a specific energy until we measure it.
This was the position of Bohr, Einstein's sparring partner in this debate,
but Einstein though that 
``[n]o reasonable definition of reality could be expected to permit this.''\footnote{A. Einstein,
B. Podolsky and N. Rosen, ``Can Quantum-Mechanical Description of Physical Reality Be Considered Complete?''
\textit{Physical Review} \textbf{47}, 777-780, 1935.}

Who was correct, Bohr or Einstein?
At the time, it was impossible to determine, since both agreed that the probabilistic interpretation
correctly predicted experimental outcomes, which it did.
Einstein thought there was something more, but couldn't prove it (except by appeal to his version of ``reality'').
Bohr thought there was nothing more, but also couldn't prove it.
It was not until the 1960s that John Bell
came up with a way to determine who was right and who was wrong.
He derived an inequality that would be violated if standard quantum mechanics was correct (that is, Bohr), 
and satisfied if Einstein's reality were correct.
It wasn't until the 1970s (long after Einstein's death in 1955 and Bohr's death in 1962)
that technology had developed to the point where experiments could be done that
were sensitive enough to distinguish between the two cases.
All experiments that have been performed up to the present come down solidly on the side of Bohr.\footnote{See, for example,
J.\ Bell,
\textit{Speakable and unspeakable in quantum mechanics:
Collected papers on quantum philosophy}
(Cambridge University Press, New York, 2004).}
There is no underlying reality to the energy of an electron.
It truly is in a superposition state, and the energy is not determined until it is measured.

\end{document}